\documentclass[preprint]{JHEP}

\usepackage{epsfig}
\usepackage{tabularx}
\usepackage{graphicx}
\usepackage{amssymb,amsfonts,amsmath}

\title{Cosmic strings and Natural Inflation}

\author{Grigoris Panotopoulos\\
ASC, Department of Physics LMU,\\
Theresienstr. 37, 80333 Munich, Germany\\
{\tt E-mail: grigoris@theorie.physik.uni-muenchen.de}}

\abstract{In the present work we discuss cosmic strings in natural inflation. Our analysis is based entirely on the CMB quadrupole temperature anisotropy and on the existing upper bound on the cosmic string tension. Our results show that the allowed range for both parameters of the inflationary model is very different from the range obtained recently if cosmic strings are formed at the same time with inflation, while if strings are formed after inflation we find that the parameters of the inflationary model are similar to the ones obtained recently.}

\begin{document}

\maketitle 

\setcounter{equation}{0}

\section{Introduction}

Inflation~\cite{Guth:1980zm} solves some outstanding problems of standard Hot Big-Bang cosmology (flatness, horizon, monopole problems) and at the same time it is a theory of the primordial cosmological fluctuations responsible for the large scale structure we observe today. Simple single-field inflationary models make some robust predictions, namely that the universe is flat and that quantum fluctuations of the inflaton field during the inflationary era generate adiabatic, gaussian density fluctuations with a nearly scale-invariant power spectrum. Today these predictions can be tested by the accumulated observational data from Supernovae~\cite{Tonry:2003zg}, BOOMERanG and MAXIMA~\cite{Ruhl:2002cz}, galaxy surveys~\cite{2d} and WMAP~\cite{Spergel} which have strengthened the inflationary idea. However, there is not a theory for inflation yet. Fundamental scalars have not been observed and the self-interaction potential for the inflaton field cannot emerge naturally from a fundamental theory in a unique way. Moreover, a general class of models involving a single slowly-rolling field requires a specially designed flat potential. This is the so-called fine-tuning problem in inflation. The model of natural inflation~\cite{Freese:1990rb} was born sixteen years ago in order to address this problem. In this model for inflation, which can be realized in certain particle physics models~\cite{Adams:1992bn}, the inflaton potential is flat due to shift symmetries. Recently the model of natural inflation has been reexamined~\cite{Freese:2004un, Savage:2006tr} in the light of recent data. The potential for the inflaton field is determined by two mass scales, $f$ and $\Lambda$. It has been shown that successful inflation as well data from Wilkinson Microwave Anisotropy Probe (WMAP)~\cite{Spergel} require $f > 0.7 m_{pl}$, where $m_{pl}=1.22 \times 10^{19}$~GeV is the Planck mass, and $\Lambda \sim 10^{15}$~GeV.

The particle physics Standard Model (SM) (for a review see e.g.~\cite{Langacker:1995hi}) has been tested to a very high precision. However, small but non-vanishing neutrino masses~\cite{Fukuda:1998ua, Ahmad:2002jz} is perhaps the most direct evidence that one should go beyond the SM. A very well-motivated theoretical framework for physics beyond the SM is the Supersymmetric Grand Unified Theory (SUSY-GUT). Supersymmetry (for a review on supersymmetry and supergravity see e.g.~\cite{Nilles:1983ge}) elegantly solves the hierarchy problem of particle physics, provides a popular candidate for Cold Dark Matter, makes the Higgs mechanism more natural compared to the SM, and the gauge coupling constants of strong, weak and electromagnetic interactions meet at a single point $M_{GUT}=2 \times 10^{16}$~GeV. GUTs~\cite{Langacker:1980js} imply a sequence of spontaneous symmetry breaking (SSB) of the GUT gauge group $G_{GUT}$ down to the SM gauge group $G_{SM}=SU(3) \times SU(2) \times U(1)$. The energy scale at which this sequence starts is $M_{GUT}=2 \times 10^{16}$~GeV. Usually GUT models predict the appearance of stable topological defects~\cite{defects}, monopoles, strings, domain walls and textures. Topological defects are produced in cosmological phase transitions by the Kibble mechanism~\cite{Kibble:1980mv}. Among the various defects, textures do not have important cosmological consequences, while monopoles and domain walls are undesirable since they lead to catastrophic cosmological implications. A common mechanism to get rid of the unwanted topological defects is to introduce cosmological inflation. On the other hand, cosmic strings could contribute to the anisotropies of cosmic microwave background (CMB). In fact, topological defects were first introduced in cosmology as a mechanism to produce the primordial perturbations needed in the early universe to form the large-scale structure we observe today. However now we know that topological defects alone cannot fit the CMB angular power spectrum~\cite{Fraisse:2006xc}.

Natural inflation does not require the formation of cosmic strings. Despite this, we shall assume that cosmic strings may be formed independently during a phase transition that is not related to the particular inflationary model. Then both inflation and cosmic strings contribute to the CMB quadrupole temperature anisotropy. We use the observational value for the quadrupole as well as known bounds on the cosmic string tension. We will show that the allowed range for the parameters of natural inflation is quite different than that obtained in a recent work if cosmic strings are formed at the same time with inflation. If however cosmic strings are formed after inflation, then the parameters of the model are similar to the ones obtained recently.

Our work is organized as follows: In the next section we present the theoretical framework. In the third section we discuss cosmic strings in natural inflation and finally we conclude in the final section.

\section{Theoretical framework}

\subsection{Natural inflation}

Here we review the natural inflation model~\cite{Freese:1990rb, Freese:2004un, Savage:2006tr}. We assume that a global symmetry is spontaneously broken at some scale $f$, with a soft explicit symmetry breaking at a lower scale $\Lambda$. According to Goldstone's theorem, the spontaneous symmetry breaking of a global symmetry leads to a Goldstone boson. If the global symmetry is an exact one, then the Goldstone scalar is exactly massless. However, if the symmetry is explicitly broken, then the scalar acquires a mass. The two scales $f$ and $\Lambda$ completely define the model. The resulting potential for the Pseudo-Nambu-Goldstone Boson (PNGB), which we choose to be our candidate for inflation, has the generic form
\begin{equation}
V(\phi)=\Lambda^4 \: (1 \pm \textrm{cos}(N \phi/f))
\end{equation}
and the mass of the PNGB is given by
\begin{equation}
m_{\phi}=\frac{\Lambda^2}{f}
\end{equation}
A very well-known example of a PNGB is the QCD axion~\cite{Weinberg:1977ma}, which solves the strong $CP$ problem by the Peccei-Quinn (PQ) mechanism~\cite{Peccei:1977hh}. In the axion case, $\Lambda$ is the QCD scale, $\Lambda \sim 200$~MeV, while $f$ is the scale at which the PQ symmetry $U(1)_{PQ}$ is spontaneously broken, typically $f \sim 10^{11}$~GeV.

Using the WMAP data~\cite{Spergel} for the amplitude $A_s$ and the spectral index $n_s$ for density perturbations we can determine the height ($\sim \Lambda$) and the width ($\sim f$) of the potential. The bound for the spectral index $n_s$
leads to a bound for the scale $f$
\begin{equation}
f > 0.7 m_{pl}
\end{equation}
while the COBE normalization for the amplitude $A_s$ determines the scale $\Lambda$
\begin{equation}
\Lambda \sim 10^{15}~GeV
\end{equation}
With $f \sim m_{pl}$ and $\Lambda \sim 10^{15}$~GeV we obtain for the inflaton mass $m_{\phi} \sim 10^{11}$~GeV.

One more thing that one should take care of is a satisfactory reheating after inflation. At the end of the slow-roll regime the inflaton field begins to oscillate about the minimum of the potential leading to entropy and particle production. If the decay rate of the inflaton is $\Gamma_{\phi}$, then the reheating temperature after inflation $T_R$ is given by
\begin{equation}
T_R=\left ( \frac{45}{4 \pi^3 g_{eff}} \right )^{1/4} \: \sqrt{\Gamma_{\phi} m_{pl}}
\end{equation}
where $g_{eff}$ is the effective number of degrees of freedom for relativistic particles at the reheating temperature. If the reheating temperature is $T_R > 1$~TeV then all particles are relativistic and for the standard model $g_{eff}=106.75$, while for the MSSM $g_{eff}=228.75$. The inflaton decay rate on dimensional grounds is given by
\begin{equation}
\Gamma_{\phi}=g^2 \: \frac{m_{\phi}^3}{f^2}=g^2 \: \frac{\Lambda^6}{f^5}
\end{equation}
where $g$ is some effective coupling constant with values $g \leq 1$. Any viable model for inflation must satisfy the gravitino constraint~\cite{Ellis:1982yb, Cyburt:2002uv} $T_R \leq (10^{6}-10^{7})$~GeV. With $f \sim m_{pl}$ and $\Lambda \sim 10^{15}$~GeV, one can easily see that the gravitino constraint is satisfied for $g \leq 1$.

\subsection{The CMB temperature anisotropies}

The temperature anisotropies~\cite{Durrer:1996nq} are usually quoted in terms of coefficients $C_l$ of two-point correlations. The anisotropy can be expanded in spherical harmonics, which is a natural basis because the last scattering surface, from which the photons of the CMB reach us today on earth, is a sphere. So we can write
\begin{equation}
\Theta(\vec{n}) \equiv \frac{\delta T}{T}(\vec{n})=\sum_{l,m} \: a_{lm} Y_l^m(\vec{n})
\end{equation}
where the unit vector $\vec{n}$ is the direction of observation in the sky, and we omit the indices $\eta_0$ for today and $x_0$ for ``here''. The two-point correlation between two directions in the sky is given by
\begin{equation}
C(\theta) \equiv \langle \Theta({\vec{n}_1})\Theta(\vec{n}_2) \rangle = \sum_{l,l',m,m'} \: \langle a_{l,m} a_{l',m'}^{*} \rangle Y_l^m(\vec{n}_1) (Y_{l'}^{m'}(\vec{n}_2))^{*}
\end{equation}
where $\theta$ is the angle between the directions $\vec{n}_1 \cdot \vec{n}_2 = cos \theta$. The coefficients $C_l$ are defined by
\begin{equation}
\langle a_{l,m} a_{l',m'} \rangle = \delta_{l l'} \delta_{m, m'} C_l
\end{equation}
Using this definition and the addition theorem for spherical harmonics we obtain
\begin{equation}
C(\theta)=\frac{1}{4 \pi} \sum_{l} \: (2l+1) C_l P_l(cos \theta)
\end{equation}
where $P_l(x)$ are the Legendre's polynomials of degree $l$. The rms value for the temperature anisotropies is defined~\cite{books}
\begin{equation}
\left ( \frac{\delta T}{T} \right )_{rms}^2 \equiv C(\theta=0)=\frac{1}{4 \pi} \sum_{l} \: (2l+1) C_l P_l(1)=\frac{1}{4 \pi} \sum_{l=2} \: (2l+1) C_l
\end{equation}
where we have neglected the first two terms corresponding to $l=0$ (monopole term) and $l=1$ (dipole term). The monopole term is unobservable, while the dipole term has to do with the relative motion of the observer with respect to the CMB frame and has been removed for the determination of the rms value of the anisotropy.
The dominant contribution to the sum above comes from the first term for $l=2$, which is the quadrupole anisotropy~\cite{books}
\begin{equation}
\left ( \frac{\delta T}{T} \right )_{Q}^2=\frac{5}{4 \pi} \: C_2
\end{equation}
and for which the observational value is~\cite{Spergel}
\begin{equation}
\left ( \frac{\delta T}{T} \right )_{Q}=6.6 \times 10^{-6}
\end{equation}

\section{Natural inflation meets cosmic strings}

For the inflaton potential we choose to work with $N=1$ and the minus sign. In the slow-roll approximation the slope and the curvature of the potential must satisfy the two constraints $\epsilon \ll 1$ and $|\eta| \ll 1$, where $\epsilon$ and $\eta$ are the two slow-roll parameters which are defined by
\begin{equation}
\epsilon \equiv -\frac{\dot{H}}{H^2}
\end{equation}
\begin{equation}
\eta \equiv \frac{V''}{3 H^2}
\end{equation}
The end of inflation is determined by the condition $max (\epsilon, |\eta|)=1$.
In this approximation the equation of motion for the scalar field takes the form
\begin{equation}
\dot{\phi} \simeq -\frac{V'}{3 H}
\end{equation}
while the Friedmann equation becomes $(V \gg \dot{\phi}^2)$
\begin{equation}
H^2 = \frac{8 \pi G}{3} \: V(\phi)
\end{equation}
where $G$ is Newton's constant.
The number of e-folds during inflation is given by
\begin{equation}
N_* \equiv \textrm{log} \frac{a_{end}}{a_{*}} = \int _{t_{*}}^{t_{end}} \: H(t) dt
\end{equation}
For a strong enough inflation so that the horizon problem is solved we take $N_* =60$.

Assuming that in the total quadrupole anisotropy contribute both inflation (infl) and cosmic strings (cs), then the anisotropy will be given by
\begin{equation}
\left ( \frac{\delta T}{T} \right )_{Q}^2=\left ( \frac{\delta T}{T} \right )_{Q-infl}^2+\left ( \frac{\delta T}{T} \right )_{Q-cs}^2
\end{equation}
The contribution from inflation has a scalar part and a tensor part. The scalar part is given by~\cite{Rocher:2004et}
\begin{equation}
\left ( \frac{\delta T}{T} \right )_{Q-scal}=\frac{1}{4 \pi \sqrt{45}} \: \frac{V^{3/2}(\phi_*)}{M_p^3 V'(\phi_*)}
\end{equation}
where $M_p=2.43 \times 10^{18}$~GeV is the reduced Planck mass, $V(\phi)$ is the potential for the inflaton field, $V'$ is the derivative $dV/d \phi$ of the potential with respect to $\phi$ and $\phi_*$ is the value of the inflaton field when the comoving scale corresponding to the quadrupole anisotropy became larger than the Hubble radius. The tensor part is given by~\cite{Rocher:2004et}
\begin{equation}
\left ( \frac{\delta T}{T} \right )_{Q-tens}=\frac{0.77}{8 \pi} \: \frac{V^{1/2}(\phi_*)}{M_p^2}
\end{equation}
Therefore the total quadrupole anisotropy is given by
\begin{equation}
\left ( \frac{\delta T}{T} \right )_{Q}^2=\left ( \frac{\delta T}{T} \right )_{Q-scal}^2+\left ( \frac{\delta T}{T} \right )_{Q-tens}^2+\left ( \frac{\delta T}{T} \right )_{Q-cs}^2
\end{equation}
where $\left ( \frac{\delta T}{T} \right )_{Q-scal}$ and $\left ( \frac{\delta T}{T} \right )_{Q-tens}$ are given above and the contribution from cosmic strings is given by~\cite{Rocher:2004et}
\begin{equation}
\left ( \frac{\delta T}{T} \right )_{Q-cs}=(9-10)G \mu
\end{equation}
where $\mu=2 \pi v^2$ and $v$ is the vacuum expectation value (vev) of the Higgs field responsible for the formation of cosmic strings.

The effects of cosmic strings depend on the value of the dimensionless quantity $G \mu$, where $\mu$ is the tension of the string (mass per unit length) and sets the scale of SSB at which cosmic strings are formed. Current observations constrain the string tension to be $G \mu < 2 \times 10^{-7}$~\cite{Pogosian}. A stronger bound comes from pulsars, $G \mu < 10^{-7}$~\cite{pulsar}, and an even stronger bound comes from reionization, $G \mu \leq 3 \times 10^{-8}$~\cite{Vilenkin}. The bound on the string tension puts an upper bound on the Higgs vev
\begin{equation}
v < 2.2 \times 10^{15}~GeV
\end{equation}
for the weakest bound on $\mu$, and
\begin{equation}
v \leq 8.4 \times 10^{14}~GeV
\end{equation}
for the strongest bound on $\mu$.

Below we shall assume that inflation takes place after monopole formation, so that the unwanted monopoles are diluted away, and that cosmic strings are formed either after inflation or at the same time with inflation. If cosmic strings are formed prior to inflation they will be diluted away as well and they will not have any observational effect on the subsequent evolution of the universe.

First we focus on the case in which cosmic strings are formed at the same time with inflation. In this case the Higgs vev $v$ for the SSB that generates the cosmic strings and the inflation scale $M_I$ are of order $\Lambda$, $M_I \sim v \sim \Lambda$, which should satisfy the upper bound written above. We use the expression for the quadrupole temperature anisotropy, which is essentially a function of the parameters of the model $f$ and $\Lambda$,  and require that it should satisfy the observational value, namely
\begin{equation}
\left ( \frac{\delta T}{T} \right )_{Q}(f,\Lambda)=6.6 \times 10^{-6}
\end{equation}
The equation above relates $f$ and $\Lambda$ in a certain fashion. Defining the dimensionless quantities
\begin{eqnarray}
f_{18} & \equiv & \frac{f}{10^{18}~GeV} \\
\Lambda_{15} & \equiv & \frac{\Lambda}{10^{15}~GeV}
\end{eqnarray}
the graph $\Lambda_{15}$ versus $f_{18}$ can be shown in the Fig. 4 below. We see that it is a continuous increasing function which saturates as $\Lambda \sim 4 \times 10^{15}$~GeV. In fact the curve that we have obtained is similar to the one obtained in~\cite{Rocher:2004et}. If then we impose the weak upper bound $\Lambda < 2.2 \times 10^{15}$~GeV, we obtain for $f$, $f < 5.8 \times 10^{18}$~GeV, while if we impose the strong bound $\Lambda \leq 8.4 \times 10^{14}$~GeV we obtain $f \leq 4.9 \times 10^{18}$~GeV. Therefore, using the strongest bound for the string tension we obtain the results
\begin{eqnarray}
\Lambda & \leq & 8.4 \times 10^{14}~GeV \\
f & \leq & 4.9 \times 10^{18}~GeV
\end{eqnarray}
This is the first of our main results in this article. It is obvious that the allowed range for both parameters of the inflationary model is very different from the range obtained recently.

Now we go on to the case in which cosmic strings are formed after inflation and we impose the strongest bound to the Higgs vev, $v \leq 8.4 \times 10^{14}$~GeV. This bound sets a condition for the inflation contribution to the quadrupole
\begin{equation}
\left ( \frac{\delta T}{T} \right )_{Q-infl}^2 \geq 4.3 \times 10^{-11}
\end{equation}
Since inflation takes place after monopole formation and prior to cosmic string formation the condition $v < \Lambda \sim M_I < M_{GUT}$ should be satisfied.
For a given $f$ we can compute the total (scalar plus tensor) inflation contribution to the quadrupole as a function of $\Lambda$. This can be seen in Fig. 1, 2 and 3 below (they are of parabolic form, that is the anisotropy in y axis $\sim \Lambda^2$) for several different values of $f$. We obtain a lower bound on $\Lambda$ as follows
\begin{itemize}
\item For $f=8.6 \times 10^{18}~GeV$, \quad $\Lambda \geq 7.6 \times 10^{15}~GeV$
\item For $f=9.5 \times 10^{18}~GeV$, \quad $\Lambda \geq 9.1 \times 10^{15}~GeV$
\item For $f=1.2 \times 10^{19}~GeV$, \quad $\Lambda \geq 1.2 \times 10^{16}~GeV$
\end{itemize}
This is the second main result in the present work. We see that the lower bound on $\Lambda$ increases with $f$. So in order to have a
scale $\Lambda$ between $v$ and $M_{GUT}$ the scale $f$ has to be close to Planck mass.
Furthermore, there is a narrow allowed range for $\Lambda$ of the order of $\Lambda \sim 10^{15}$~GeV. These values for the inflationary parameters $f, \Lambda$, are similar to the ones obtained recently.

\section{Conclusions}

In the present work we have discussed cosmic strings in natural inflation. Cosmic strings are topological defects that generally appear in SSB schemes of GUT models. Measurements of the CMB temperature anisotropies constrain the contribution of the cosmic strings to the angular power spectrum of the anisotropies, and in general there are several works that impose an upper bound (weaker or stronger) on the cosmic string tension. On the other hand, natural inflation was invented in order to address the so-called fine-tuning problem in inflation. The model is completely characterized by two mass scales, $f$ and $\Lambda$. The scale $f$ is the scale at which a global symmetry is spontaneously broken and determines the width of the potential for the resulting PNGB, while the lower scale $\Lambda$ is related to a soft explicit symmetry breaking and determines the height of the potential. The recent data from WMAP determine the parameters of the model. According to a recent work one obtains $f \sim m_{pl}=1.22 \times 10^{19}$~GeV and $\Lambda \sim 10^{15}$~GeV. The present work is based entirely on the CMB quadrupole temperature anisotropy and on the existing upper bounds on the cosmic string tension $\mu$. Our work shows that if cosmic strings are formed at the same time with inflation, the observed value for the quadrupole anisotropy relates the scales $f$ and $\Lambda$ in a certain fashion. Then the upper bound on $\mu$ induces an upper bound on both $\Lambda$ and $f$. In particular, our analysis show that $f < 5.8 \times 10^{18}$~GeV for the weakest bound and $f \leq 4.9 \times 10^{18}$~GeV for the strongest bound. We conclude that our results are different compared to that obtained in previous works. However, if cosmic strings are formed after inflation, we find that for $f \sim m_{pl}$ there is a narrow allowed range for $\Lambda$ of the order $\Lambda \sim 10^{15}$~GeV, which is of the same order of magnitude obtained recently.

\section*{Acknowlegements}

We would like to thank T.~N.~Tomaras for reading the manuscript and for useful discussions. The present work was supported by project "Particle Cosmology".

\newpage

\begin{figure}
\centerline{\epsfig{figure=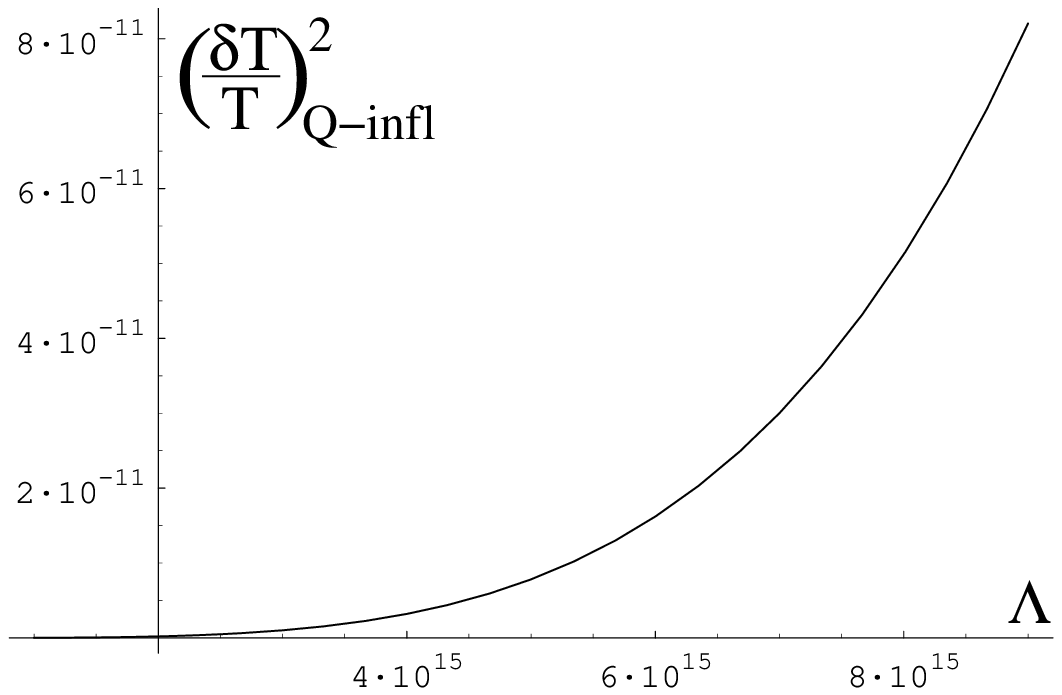,height=6cm,angle=0}}
\caption{Inflation contribution to the quadrupole anisotropy as a function of $\Lambda$ (in GeV)
for $f=8.6 \times 10^{18}~GeV$.}
\end{figure}

\begin{figure}
\centerline{\epsfig{figure=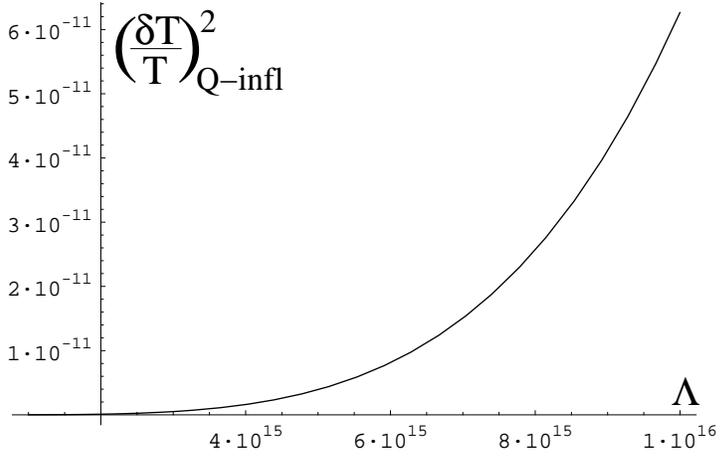,height=6cm,angle=0}}
\caption{Same as Fig. 1 but
for $f=9.5 \times 10^{18}~GeV$.}
\end{figure}

\begin{figure}
\centerline{\epsfig{figure=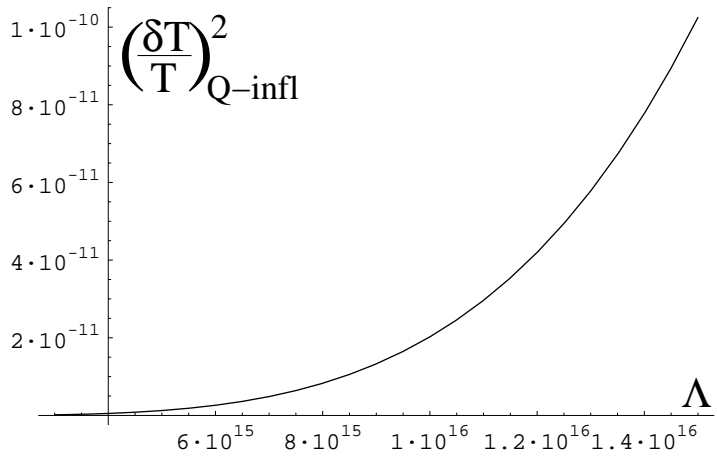,height=6cm,angle=0}}
\caption{Same as Fig. 1 but
for $f=1.2 \times 10^{19}~GeV$.}
\end{figure}

\begin{figure}
\centerline{\epsfig{figure=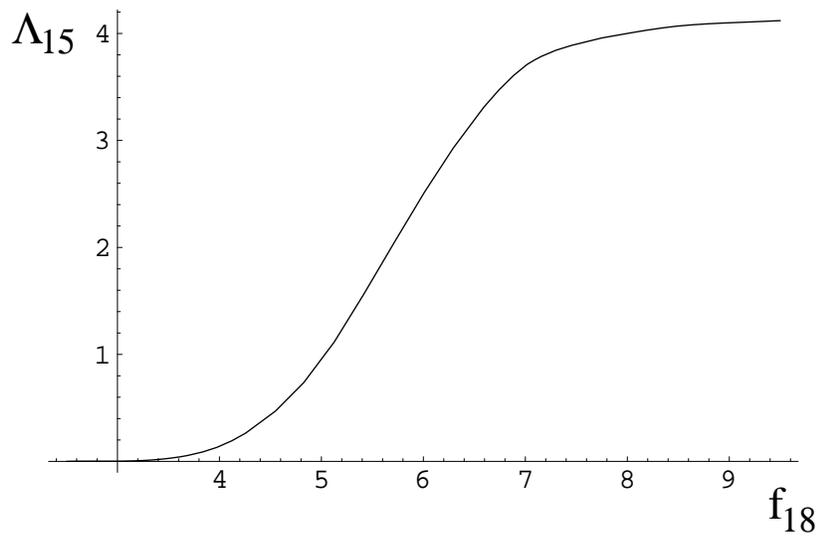,height=7cm,angle=0}}
\caption{Shown is $\Lambda_{15}$ versus $f_{18}$. The quantities $\Lambda_{15}$ and $f_{18}$ are defined in the text.}
\end{figure}

\end{document}